\documentstyle[multicol,epsf,aps]{revtex}
\begin{document}
\draft
\title{Crossover from directed percolation to compact directed percolation}

\author{J. F. F. Mendes$^{1,(a)}$, R. Dickman$^{2,(b)}$ and 
H. Herrmann$^{3,4,(c)}$}

\address{$^1$Center for Polymer Studies and Physics Department,
Boston University, Boston, MA 02215
and Department of Physics, University of Porto,
Rua do Campo Alegre 687, 4150 Porto -- Portugal\\
$^2$Department of Physics and Astonomy, Herbert H. Lehman College,
City University of New York, Bronx, New York 10468\\
$^3$P.M.M.H. (U.R.A. 857), E.S.P.C.I. Paris, 10 rue Vauquelin, 73231 Paris,
France\\
$^4$ICA 1, University of Stuttgart, Pfaffenwaldring 27, 70569
Stuttgart, Germany
}

\maketitle

\begin{abstract}
We study critical spreading in a surface-modified directed 
percolation model in which the
left- and right-most sites have different occupation probabilities
than in the bulk.  As we vary the probability for growth at an edge,
the critical exponents switch from the
compact directed percolation class to ordinary directed
percolation.  We conclude that the
nonuniversality observed in models with multiple absorbing
configurations cannot be explained as a simple surface effect. 

\end{abstract}

\pacs{PACS numbers: 05.70.Jk, 05.50.+q, 64.60.-i, 02.50.-r }



\begin{multicols}{2}

\narrowtext

Recently, considerable effort has been devoted to understanding phase
transitions in nonequilibrium systems.
Many studies of models with a continuous 
transition to a {\it unique} absorbing state have established
that such
transitions belong generically to the class of directed
percolation (DP) \cite{varios}, as predicted by
Janssen \cite{jan81}, Grassberger \cite{gra82}, and
(for multicomponent models) by Grinstein {\it et al.} \cite{gri89}.
DP has been of great interest
since its introduction by Broadbent and Hammersley \cite{broad57},
and is relevant to a vast
range of models in 
biology, chemistry and physics \cite{nicolis,haken}.
Examples are catalytic reactions on surfaces \cite{ziff,gri89},
epidemics \cite{harris74}, transport in porous media \cite{bell83},
chemical reactions \cite{sch72,gratorre}, self-organized criticality
\cite{obu90,pac94},
electric current in a diluted diode network \cite{red81}, Reggeon field theory
\cite{gribov,brower,gra-sun}, and more recently in 
damage spreading \cite{gra95},
and models of growing surfaces with a roughening transition \cite{alon}.

Unlike systems with a unique absorbing configuration, understanding of
models with multiple absorbing configurations 
\cite{albano,ditri,iwan93,mend94,dick96} is far
from complete. In such models 
the critical exponents that govern spreading from a seed
vary continuously with the particle density in the initial configuration,
and obey a generalized hyperscaling relation \cite{mend94}.
The exponents assume the usual DP
values only for initial configurations having the ``natural" 
particle density ---
that of absorbing configurations generated by the system, running at the
critical point.  Nonuniversality in these models remains a puzzle.
In searching for an explanation, and remarking that
the exponents depend on the environment into which the population
spreads, one is led to investigate whether nonuniversality is 
also produced by modifying the process at the {\em surface}
of the active region.  In this work we consider surface-modified DP
in one spatial dimension, and find it
does not show continuously-variable exponents, but rather
a crossover between two 
different universality classes, compact and standard DP.
Besides providing a simple example connecting the two classes, our study
shows that the nonuniversality observed in systems
with multiple absorbing states is not simply a surface effect.

In bond directed percolation on the square lattice, 
bonds connect each site $(x,t)$ with $(x,t-1)$ and $(x-1,t-1)$.
Each bond is ``wet" with probability $p$; with
probability $1-p$ a bond is ``dry".
Suppose the origin is the only wet site in the layer $t=0$.  One 
constructs the 
cluster ${\cal C}_0$ connected to the origin using the rule that  
$(x,t) \in {\cal C}_0$ iff it is connected, by a wet bond, to  
$(x,t-1)$ or $(x-1,t-1) \in {\cal C}_0 $.
For $p<p_c$ such clusters are finite with probability 1; for
$p>p_c$ there is a nonzero probability of a cluster growing
indefinitely.  The percolation threshold, $p_c$, marks a continuous
phase transition.  The model is readily generalized by introducing
conditional probabilities $P(1|1,0) = P(1|0,1) = p_1$ and $P(1|1,1) = p_2$
for site $(x,t)$ to be wet (=1), given the states of $(x,t-1)$ and
$(x-1,t-1)$.  ($P(1|0,0)=0$, naturally.)  The resulting Domany-Kinzel
model exhibits a line of critical points in the DP class \cite{dom84}.
The endpoint of this line, $(p_1=1/2,p_2=1)$, describes a transition
outside the DP class; it corresponds instead to {\em compact directed
percolation} (CDP) \cite{essam89}.
The essential difference between DP and CDP is that in the latter,
transitions from wet to dry cannot occur within a string of 1's;
the evolution of a string of 1's is governed by a pair of random
walks at its ends.

\begin{figure}
\epsffile{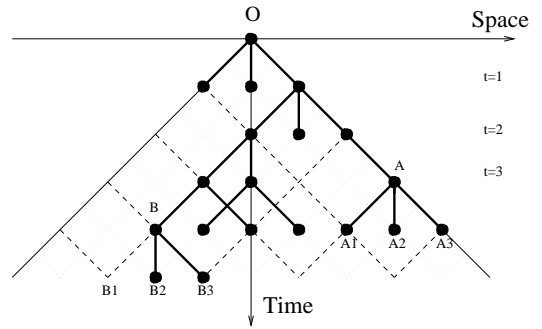}
\caption{The {\it centered square lattice} and a growing cluster.}
\end{figure}

In the present work we study directed percolation on the
{\em centered square lattice}, in which $(x,t)$ is connected
to $(x-1,t-1)$, $(x,t-1)$, and $(x+1,t-1)$ by bonds that,
as before, are wet with probability $p$ (see Fig. 1).
Earlier studies of this model \cite{tang92,duarte93},
yielded a percolation threshold of $p_c =0.5387\pm0.0003$.
We now modify the probabilities for introducing
wet sites at the surface of the active region.  Suppose $(x_R,t)$
is the rightmost wet site in layer $t$.  We set the probability for
$(x_R+1,t+1)$ to be wet at $p'$ rather than $p$, and similarly for
$(x_L-1,t+1)$, if $(x_L,t)$ is the leftmost wet site in layer $t$.
In Fig. 1, for example,
the probabilities to introduce wet
sites at
$A_1$, $A_2$ and $A_3$, are  $p$, $p$ and $p'$, respectively.
(Similarly, we have probabilities $p'$, $p$ and $p$
for introducing wet sites at
$B_1$, $B_2$ and $B_3$ .)
 
We used {\it time-dependent} simulations to study the critical behavior. 
The method involves starting with a single  wet site at $t=0$
and following the spread of wet sites in a large set of 
independent realizations.
We measured the survival probability $P(t)$ (the probability that there is 
at least one wet site at time $t$), the average number of
wet sites, $\bar{n}(t)$, and the average mean square distance of spreading
from the origin, $\bar{R}^{2}(t)$.
At criticality, these quantities follow power laws in the 
long-time limit, e.g., $P(t) \sim t^{-\delta}$,
and similarly for the other quantities.
In the subcritical phase ($p<p_{c}(p')$)
the survival probability decays exponentially, while in the
supercritical phase ($p>p_{c}(p')$) it approaches
$P_{\infty} \equiv \lim_{t\rightarrow \infty} P(t) >0$, the {\it
ultimate survival probability}.
If we allow the exponents to depend on the surface
probability, $p'$, the simplest generalization
of the usual scaling hypothesis \cite{gratorre} is
\begin{eqnarray}
P(t) \sim t^{-\delta(p')} \Phi(\Delta t^{1/\nu_{\|}(p')}) \\
\bar{n}(t) \sim t^{\eta(p')} \Theta(\Delta t^{1/\nu_{\|}(p')})\\
\bar{R}^{2}(t) \sim t^{z(p')}\Omega(\Delta t^{1/\nu_{\|}(p')})
\end{eqnarray}
where $\Delta=|p-p_{c}(p')|$ measures
the distance to the critical point and
$\nu_{\|}(p')$ is the critical exponent associated with
the correlation length
in the time direction ($\xi_{\|} \sim \Delta^{-\nu_{\|}(p')}$).
Assuming that
the scaling functions $\Phi$, $\Theta$ and $\Omega$ are
nonsingular
at the critical point, it follows
\begin{figure}[h]
\epsfxsize=64mm
\epsffile{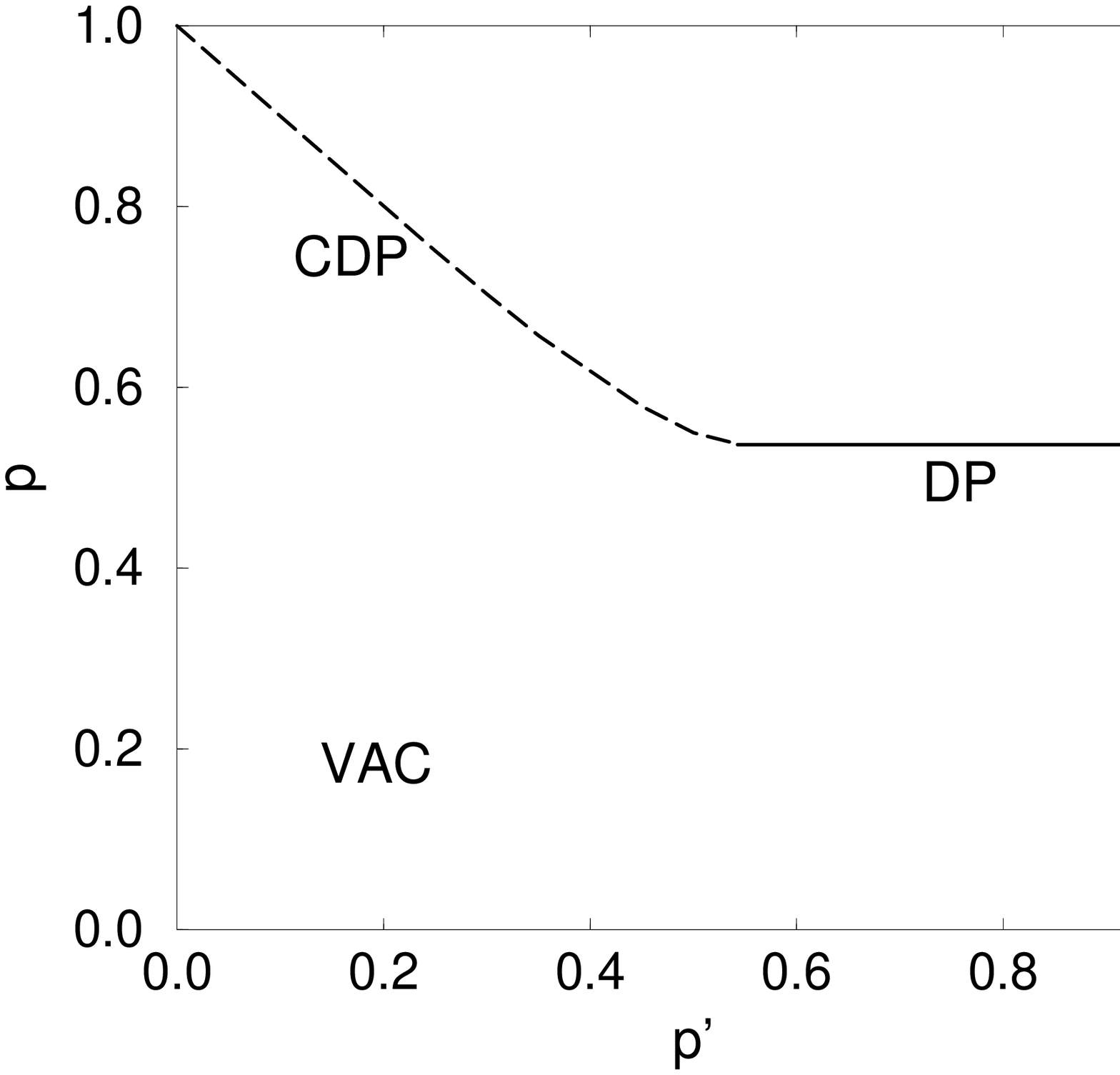}
\caption{Phase diagram in $p$-$p'$ plane. The two regimes
(CDP and DP) are shown.}
\end{figure}

{{TABLE 1.} Values of the critical exponents and
threshold percolation for different values of $p'$. The values for DP were
taken from \cite{gri89}, \cite{cardy80}, \cite{iwan93} and \cite{bax88}. The
numbers in parentheses represent the uncertainty in the last digit(s).}
\begin{tabular}{cccccc}
\hline \hline
 $p'$      & $p_c$      & $\delta$  & $\eta$    &   $z/2$      & $\beta'$ \\
\hline
 DP        & -          & 0.1596(4) & 0.312(2)  & 0.6321(4) & 0.2767(4)  \\
 CDP       & -            & 1/2       & 0         & 1         & 1      \\
\hline
0.0        & 1.0          &   -       &    -      &    -      & -      \\
0.10       & 0.9005(3)    & 0.495(7)  & 0.016(14) & 0.502(8)  & -       \\
0.25       & 0.7510(4)    & 0.496(5)  & 0.015(13) & 0.505(7)  & 0.97(4) \\
0.30       & 0.7035(5)    & 0.498(8)  & 0.022(15) & 0.509(13) & -         \\
0.35       & 0.6575(5)    & 0.493(4)  & 0.018(13) & 0.505(11) & 0.95(8) \\
0.45       & 0.5792(5)    & 0.497(7)  & 0.013(16) & 0.504(6)  & -       \\
0.50       & 0.55012(4)   & 0.499(5)  & 0.011(13) & 0.506(8)  & -       \\
$p_c$     & 0.53875(1)   & 0.159(1)  & 0.310(3)  & 0.632(2)  & 0.28(2) \\
\hline \hline
\vspace{1em}
\end{tabular}\\
that $P(t)$, $\bar{n}(t)$, and
$\bar{R}^{2}(t)$
follow pure power laws as $t \rightarrow \infty$.
A log-log plot of $P$, $\bar{n}$, or $\bar{R^2}$ as a function of time should
yield a straight line at $p=p_{c}(p')$, 
permitting one to determine $p_{c}(p')$
rather precisely.
The exponents $\delta$, $\eta$, and $z$ are given by the asymptotic
slopes of the corresponding plots.
In addition, the exponent $\beta'$ controlling the approach
to the critical point of the survival probability, 
$P_{\infty}(p) \sim (p-p_{c}(p'))^{\beta'(p')}$ \cite{beta}
may also depend on $p'$.
Our results (see Fig. 2) show
that the percolation threshold
$p_c$ increases with decreasing $p'$ for $p'<p_c^{DP}=0.53875$.
This decrease in $p'$ means that the percolating cone becomes narrow, and 
a larger  value of $p$ is needed for the cluster to survive. 

Table 1 show how the exponents vary with $p'$.
The mean number of wet sites in surviving trials is $N(t)=\bar{n}(t)/P(t)$,
and the fractal dimension $d_f$ of aggregates surviving to time $t$
is defined through $N(t) \sim R^{d_{f}}$, yielding
$d_{f}=2(\delta+\eta)/z$. For DP $d_{f}=0.752$.
In the present study
$d_f \approx 1$ for $p' < p_c^{DP}$, which means that for this range
of $p'$ the clusters are no longer fractal but {\it compact}.
\begin{figure}
\epsfxsize=85mm
\epsffile{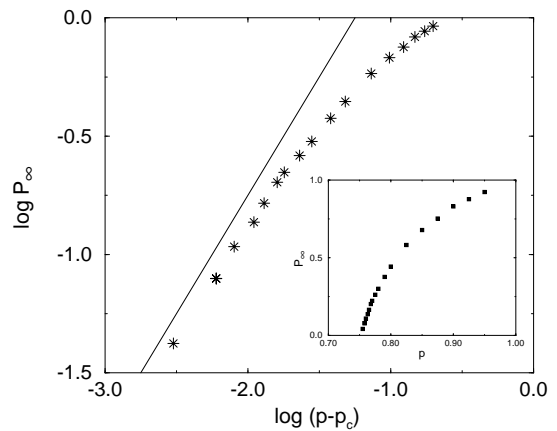}
\caption{The ultimate survival probability, $P_{\infty}$ as a function of
$p-p_{c}(p')$ in a log-log plot. The inset is a plot of $P_{\infty}$ versus
$p$. These results were obtained for $p'=0.25$ and the slope of the line is
$\beta=1$.}
\end{figure}
\begin{figure}
\epsfxsize=145mm
\epsffile{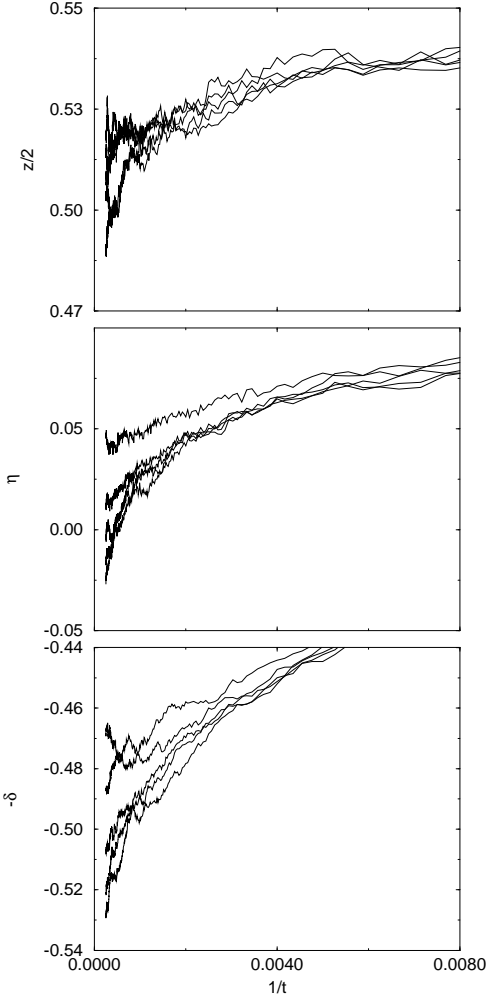}
\caption{The local slopes $-\delta(t)$ (upper), $\eta(t)$ (middle),
and $z(t)/2$ (bottom), for the case $p'=0.25$.
Each picture has five curves corresponding from bottom to top $p=0.7506$,
$0.7508$, $0.7510$, $0.7514$ and $0.7516$.}
\end{figure}
\vspace{1em}
From Fig. 3 we see that the exponent
controling the survival probability ($\beta'=0.97\pm 0.04$) is very
different from that of DP ($\beta' = \beta=0.2767\pm 0.0004$).
For $p' < p_c^{DP}$, we find $\beta = 0$, so that the generalized 
hyperscaling relation,
$(1 + \beta/\beta')\delta + \eta = dz/2$, \cite{mend94}, becomes
$\delta + \eta = dz/2$,  which describes compact growth \cite{dick95}.
The data are indeed consistent with compact hyperscaling.
(For the point nearest
$p_c$, $P_{\infty}$ is obtained from data for $t>15000$, averaged 
over $10^{5}$ independent runs.)
Figure 4 shows the local slopes as a function of inverse
time for 
$p'=0.25$. Similar curves were obtained for the other values
of $p'$. (These results reflect averages over $10^{5}$ ---
$5 \times 10^{5}$ independent trials, running to a maximum time of
$2000$ to $6000$.)
We also checked whether the scaling relations, Eqs (1) - (3), are obeyed, by
varying $\nu_{||}$ to optimize the data collapse.  For $p' < p_c^{DP}$, 
we observe scaling for $\nu_{||} = 2$, as expected for CDP.
\begin{figure}[t]
\epsfxsize=96mm
\caption{Typical evolutions from a single wet
site for $p=p'=0.45$, $p=p'=0.5387$ and $p=0.9$,
$p'=0.1$ respectively.  Time increases downward.}
\end{figure}

Our numerical results indicate
that the model exhibits a crossover from
directed percolation to compact directed percolation
as the surface growth rate, $p'$, is varied.  
We can understand this by examining
a surface-modified contact process (CP) \cite{harris74}. 
The CP (in $d$ space dimensions) 
is a sequentially-updated version of DP (in $d+1$ dimensions). In the CP, 
each lattice site is either vacant or occupied
by a particle.  
Particles die at unit rate, and give birth at rate $\lambda$.
(In a birth event a new particle appears at one of the sites neighboring
the parent, if that site is vacant.)  The CP exhibits a phase transition
in the DP class, as $\lambda $ is increased through $\lambda_c$. 

Now consider the one-dimensional CP,
 with $x_R$ and $x_L$ denoting the positions
of the right- and left-most particles at any instant, and modify the rules
so that the rate for a particle to appear at $x_R +1$, or at $x_L - 1$,
is $\lambda'/2$.
If we let $L = <x_R - x_L>$, the mean being taken over all surviving 
trials at time t, starting from a single particle, 
then $dL/dt = \lambda' - 2g$, where g is
the mean distance from $x_L$ to the next occupied site (and similarly
at the right edge).
If we set $\lambda' = \lambda$, there is a critical point at $\lambda_c$,
in which case the interior density approaches zero as $t \rightarrow \infty$.
Since $L \sim t^{z/2}$ at the critical point, $dL/dt \rightarrow 0$
as $t \rightarrow \infty$ as well.  (For $\lambda < \lambda_c$, $dL/dt < 0$,
and vice-versa.)  Hence, for $\lambda' = \lambda = \lambda_c$, $g$ must
tend to $\lambda_c/2$ at long times.  
The ultimate survival probability $P_{\infty}$
can only be nonzero if $dL/dt > 0$ at any finite time.  On the other hand,
$g$ is a decreasing function of $\lambda$ and $\lambda'$.  It follows that
if $\lambda' < \lambda = \lambda_c$, the process must die.  We can
compensate the reduced surface birth rate by increasing $\lambda$, and
should again have survival for $g(\lambda, \lambda') \leq \lambda'/2$.
Having forced $\lambda > \lambda_c$, surviving trials are compact, i.e.,
have a nonzero particle density as $t \rightarrow \infty$.
The region between $x_L$ and $x_R$ is not simply a string of
occupied sites, as it would be in CDP, but the scaling behavior should
nonetheless be characteristic of compact growth.  
When the surface growth rate {\em exceeds} the bulk critical value,
we expect a continuous (DP-like) transition at $\lambda = \lambda_c$.
For $\lambda'> \lambda_c$, the advance of the 
edges may be enhanced at
short times, but the process cannot survive for $\lambda < \lambda_c$
because the interior approaches the vacuum.  The density in surviving clusters 
is localized at the edges, so 
the number of particles remains ${\cal O}(1)$, and the process dies
with probability 1.  To summarize, our argument 
shows why surface-modified DP exhibits compact growth for
$p \geq p_c(p')$ when $p' < p_c^{DP}$, and why there is a DP-like
transition at $p_c^{DP}$ for $p' \geq p_c^{DP}$.
As in CDP, the transition is {\em first order} for $p' < p_c^{DP}$.

Clearly, the surface probability $p'$ does not play the same
role as the initial density in models with multiple absorbing configurations.
Conversely, the continuously-variable exponents observed in the latter
class of models cannot be explained as a simple surface effect.
Finally, it is interesting to note that while the surface modification
explored in the present work represents a nonlocal {\em spatial} interaction 
($x_R$ and $x_L$
are defined globally), the dynamics of models with multiple absorbing
configurations involves long {\em memory} effects \cite{munoz96}.
\vspace{1em}
\noindent

{\em Acknowledgments:}
JFFM gratefully acknowledges support from Funda\c c\~ao Luso Americana 
para o Desenvolvimento and 
PRAXIS XXI/BPD/6084/95. We thank S.~Redner and
J.~A.~M.~Duarte for helpful discussions. \\
\vspace{1em}
$^{(a)}$ Electronic address: jfmendes@sid3.bu.edu\\
$^{(b)}$ Electronic address: dickman@lcvax.lehman.cuny.edu\\
$^{(c)}$ Electronic address: hans@pmmh.espci.fr

\end{multicols}

\end{document}